# Electromagnetic modeling of large-scale high-temperature superconductor systems

Edgar Berrospe-Juarez[1], Frederic Trillaud[2], Víctor M R Zermeño[3], and Francesco Grilli[4]

[1]Postgraduate School of Engineering, National Autonomous University of Mexico, Mexico

[2]Institute of Engineering, National Autonomous University of Mexico, Mexico

[3]NKT

[4]Karlsruhe Institute of Technology, Germany

email: eberrospe@gmail.com

This work was supported in part by the Programa de Maestría y Doctorado en Ingeniería of the Universidad National Autónoma de México (UNAM) and the Consejo Nacional de Ciencia y Tecnología (CONACYT) under CVU: 490544, and by DGAPA-UNAM grant, PAPIIT-2019 IN107119.

## Abstract

The development of the high-temperature superconductors (HTS) conductors has allowed the development of diverse superconductor devices. Some of these devises, like the power generators and high-field magnets, are classified as large-scale HTS systems, because they are made of hundreds or thousands of turns. Mathematical models are required to address the analysis of these kind of systems. This task cannot be done by means of analytical models, because they are limited to the analysis of simple assemblies. The finite-element models using the H formulation have been extensively used during the last years. Nevertheless, the use of H formulation models to analyze large-scale systems is hindered by the excessive computational load. The recently proposed T-A formulation models have allowed building more efficient models for systems made of HTS tapes. Additionally, the homogenization and multi-scaling methods have been successfully applied in conjunction with the H and T-A formulations, these simplification methods allows reducing the required computational resources. In this article a new simplification method, called densification, is proposed. The strategies emerging from the combined use of the formulations and the simplification methods already mentioned are extensively explored, and the comprehensive validation and comparison of all the resulting strategies is presented.

# 1. Introduction

More than three decades after the discovery of the first high temperature superconductor (HTS) with a critical temperature above the boiling point of nitrogen, the technology has matured and the second-generation of HTS conductors (2G HTS) are commercially available [1]–[3]. The 2G HTS conductors are layered composites with a thin layer of HTS material. The 2G HTS conductors are also called 2G HTS wires, coated conductors, (RE)BCO tapes, etc. For simplicity in this article the term HTS tapes is preferred.

The emergence of HTS tapes has favored the development of diverse superconductor devices. For power systems, it is expected that cables and fault current limiters will soon reach market maturity [4]. Continuous research and development has targeted other power devices, such as transformers, generators, and superconducting magnetic energy storage systems [5], [6]. For scientific and medical applications, the interest in the technology has spawned over MRI and NMR magnets [7], [8] and high magnetic field magnets [9]. These devices are typically made of hundreds or even thousands of turns of HTS tapes, then they are classified as large-scale HTS systems [10]–[12].

To ensure safe operation, the design HTS devices must consider effects that arise from changes in external magnetic field and transport current. During these changes, hysteresis losses are generated in the HTS materials, which in extreme cases leads to the loss of the superconducting state [12]–[14]. For simplicity in this article, the hysteresis losses are simply refered as losses. The estimation of current density, electric and magnetic fields inside the superconductor is a mandatory step for obtaining the losses, as well as other quantities of interest for practical applications [7], [15]–[17]. The available analytical models are restricted to the analysis of individual tapes or relatively simple assemblies under restrictive conditions [18]–[24]. The analysis of systems with more intricate geometries and operating conditions, i.e., real HTS systems, requires the use of numerical methods [13], [25]–[31].

The finite element method (FEM) is well documented in the literature [32]–[34] and has been extensively used to address the analysis of HTS systems. The Maxwell's equations can be written using different formulations. The formulations differ from each other in the selection of the state variables. As stated in [35]: "in principle, all these formulations are equivalent, but the solutions of the corresponding partial differential equations (PDE) by means of FEM can be very different". The most frequently used formulations within the superconductor community, are: *A-V* formulation [15], [36]–[38], *T-Ω* formulation [39], [40], and *H* formulation [41], [42].

The *H* formulation as used nowadays was introduced in [41], [42]. This formulation has been widely used during the last years and has arguably become the *de facto* standard within the applied superconductivity community. A recently published review [43] claims: "at the time of writing, the *H* formulation has been used by more than 45 research groups worldwide". Nevertheless, the use of the *H* formulation models that consider in detail each individual turn/tape of the large-scale system becomes a prohibitive task in terms of computational load. This kind of models are known as *H* full models. The *T-A* formulation was recently proposed in [44], [45]. This formulation allows building more efficient models than those based on the *H* formulation. In the *T-A* formulation, the HTS tapes are modelled as infinitely thin lines, therefore the mesh complexity and the computation time can be reduced.

The limitations of the full models have favored the emergence of *simplification methods* like the homogenization and multi-scaling methods. The homogenization assumes that a stack made of HTS tapes can be represented by a single anysothropic homogeneous bulk [14]. The multi-scaling method is based on the analyzis of a reduced set of tapes, called analyzed tapes, and the subsequent approximation of the behavior of the full system [12]. As of today, these two simplification methods have been succesfuly used together with the *H* and *T-A* formulation, giving rise to the following strategies: *H* homogeneous [14], *H* multi-scale [12], *H* iterative multi-scale [46], *T-A* homogeneous [47], and *T-A* simultaneous multi-scale [47]. In the reference [47] the *T-A* simultaneous multi-scale strategy is called just *T-A* multi-scale, here the adjective "simultaneous" is added to differentiate this strategy from the multi-scale and iterative multi-scale strategies.

The first contribution of this article is the proposition of new strategies. A new simplification method, called *densification*, is proposed, thereby giving rise to the *H* densified and *T-A* densified strategies. The densification method consists in merging the HTS tapes forming part of a stack with their neighboring tapes, so that the original stack can be modeled by means of fewer tapes. The ideas of the multi-scaling method are revisited and the *H* simultaneous multi-scale strategy is proposed. The *H and T-A* simultaneous multi-scale strategies are enhanced by means of the homogenization and densification of the non-analyzed tapes, resulting in four additional strategies. Figure 1 shows a tree diagram with the different strategies that emerge from the combination of the *H* and *T-A* formulations, and the simplification methods. The blue rectangles represent the strategies already described in the literature, while the green rectangles stand for the strategies that are original contributions of this article. The second contribution of this article is the comprenhensive comparison of the strategies showed in Figure 1.

The models presented in this article were implemented in COMSOL Multiphysics 5.3 [48]. The computer used to perform the simulations is an Apple MacBookPro (3 GHz Intel Core i7-4578U, 4 cores, 16 GB of RAM). The characteristics of the computer are important to weight the reported computation times.

This article is organized as follows. Section 2 contains a brief the description of the *H* and *T-A* formulations. The case study used to compare the strategies is presented in Section 3, the reference and full models are also presented there. The rest of the strategies based on the *H* and *T-A* formulations are compiled in Sections 4 and 5, respectively. Section 6 contains the comparison and discussion of the different strategies. Finally, the conclusions are presented in Section 7.

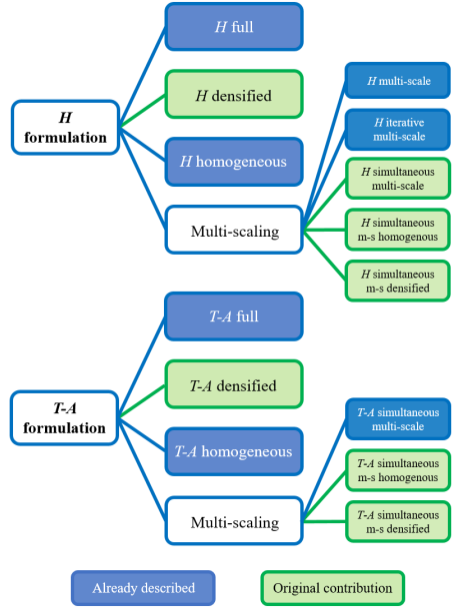

**Figure 1.** Tree diagram showing the strategies that emerge from the coupling of the formulations: *H* and *T-A*; and the simplification methods: homogenization, densification, and multi-scaling.

## 2. Formulations

In this section, we briefly recall salient information of the *H* and *T-A* formulations. For further information related to the *H* formulation, the reader is referred to [41], [43]. And, for further information related to the *T-A* formulation, the reader is referred to [44], [45].

### 2.1. *H* formulation

The *H* formulation uses the magnetic field strength **H** as dependent variable. Within a bounded universe the different materials are represented by different subdomains. Each subdomain has different properties, i.e., resistivity $\rho$ and permeability $\mu$. These values define the constitutive relations $\mathbf{E} = \rho \mathbf{J}$ and $\mathbf{B} = \mu \mathbf{H}$; where **E** and **H** are the electric and magnetic field strength, respectively; and **B** and **J** are the magnetic flux and current density, respectively.

To derive the governing equation of the *H* formulation, Ampère's law is written neglecting the displacement current. Then, Ampère's and Faraday's laws are given by,

$$\nabla \times \mathbf{H} = \mathbf{J}, \tag{1}$$

$$\nabla \times \mathbf{E} = -\frac{\partial \mathbf{B}}{\partial t}, \tag{2}$$

substituting the constitutive relations of the materials and (1) into (2) yields the governing equation,

$$\nabla \times (\rho \nabla \times \mathbf{H}) = -\mu \frac{\partial \mathbf{H}}{\partial t}. \qquad (3)$$

Gauss's law $\nabla \cdot \mathbf{B} = 0$ is fulfilled by means of the election of the initial conditions, as explained in [14], [41].

Considering a two-dimensional (2D) planar model, like the one depicted in Figure 2, the x-y plane contains the cross-section of the superconductors, and the subdomains $\Omega_{sc}$, $\Omega_n$, and $\Omega_{sm}$ represent the superconductor, normal conductor and surrounding mediums, respectively. The surrounding medium subdomain includes the insulating materials and the cryogenic liquid. In the 2D planar model, **H** has two non-zero components, while **E** and **J** have just one non-zero component each one. Therefore, restricting the study to linear magnetic materials ($\mu = const$), the governing equation (3) can be written as follows,

$$\mu \frac{\partial H_x}{\partial t} + \frac{\partial (E_z)}{\partial y} = 0, \qquad (4)$$

$$\mu \frac{\partial H_y}{\partial t} - \frac{\partial E_z}{\partial x} = 0, \qquad (5)$$

where $E_z = \rho J_z$, and

$$J_z = \frac{\partial H_y}{\partial x} - \frac{\partial H_x}{\partial y}. \qquad (6)$$

The transport currents $I_k$ in each conductor are imposed by means of integral constraints. One constraint is required for each conductor, as explained in [14], [41].

The selection of the elements used in the FEM discretization plays also an important role on the accuracy and computational speed of the numerical model. In the case of the $H$ formulation, several arguments are presented in [14], [41], [49] showing the advantages of the first-order edge elements over other kind of elements.

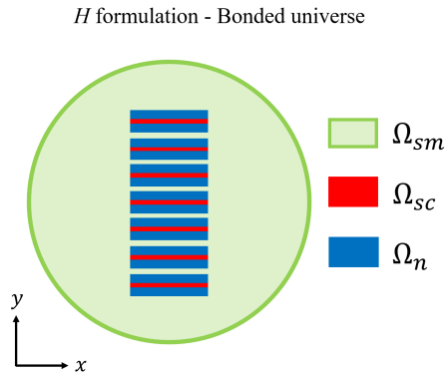

**Figure 2.** Bounded universe of the a $H$ formulation 2D planar model, formed by the union of the superconductor $\Omega_{sc}$, normal conductor $\Omega_n$, and surrounding medium $\Omega_{sm}$ subdomains.

## 2.2. *T-A* formulation

The *T-A* formulation, as described in [44], [45], relies on the primary assumption that thin superconducting layers of the HTS tapes can be modelled as one dimensional (1D) objects in a 2D model. The infinitely thin approximation is meaningful when dealing with superconductors wires having large aspect ratio (width/thickness), like the 2G HTS tapes, where this ratio is in the range of $10^4$ [1]. The *T-A* formulation requires the implementation of both the *T* and the *A* formulations, and both state variables **T** and **A**, current and magnetic vector potentials, are evaluated.

Here, a 2D planar geometry is assumed, see Figure 3. The bounded universe is made of 1D superconducting layers and the surrounding medium $\Omega_{sm}$, the normal conductor layers of the HTS tapes are not considered. It is assumed that the current only flows through the superconducting layers, and the surrounding medium is considered to be non-conductive. The current vector potential **T** is exclusively defined over the superconducting layers, while the magnetic vector potential **A** is defined over the entire bounded universe.

Considering Faraday's law (2) and the definition of the current vector potential **T** (**J** = $\nabla \times$ **T**), the governing equation of the *T* formulation is,

$$\nabla \times \rho \nabla \times \mathbf{T} = -\frac{\partial \mathbf{B}}{\partial t}. \tag{7}$$

In 2D cases, as long as the thickness of the superconducting layer is neglected, **J** and **T** have only one non-zero component, and (7) is simplified as follows,

$$\frac{\partial}{\partial x}\left(\rho \frac{\partial T_y}{\partial x}\right) = \frac{\partial B_y}{\partial t}, \tag{8}$$

The transport currents in each tape $I_k$ is imposed by modifying the boundary conditions for $T_y$. The values of $T_y$ at the edges of the 1D layer ($T_1$ and $T_2$) must fulfil the following relation,

$$I_k = (T_1 - T_2)\delta, \tag{9}$$

where $\delta$ is the real thickness of the HTS layer.

The component of **B** perpendicular to the superconducting layer $B_y$, required to compute $T_y$ in (8), is obtained by calculating **A**. Considering Ampère's law (1), and the definition of the magnetic vector potential **A** (**B** = $\nabla \times$ **A**), the governing equation of the *A* formulation is,

$$\nabla \times \nabla \times \mathbf{A} = \mu \mathbf{J}. \tag{10}$$

In 2D cases, $A_z$ is the only non-zero component of **A**, therefore (10) is simplified to,

$$\nabla^2 A_z = 0. \qquad (11)$$

At first glance, equation (10) should be simplified to $\nabla^2 A_z = -\mu J_z$, but for the purpose of computing $A_z$, $J_z$ is equal to zero all over the bounded universe. The reason is that the current flows only through the 1D superconducting layers.

In order to couple $J_z = \partial T_y/\partial x$ with the $A$ formulation, the surface current density $\mathbf{K}$ is defined as,

$$\mathbf{K} = \delta \mathbf{J}, \qquad (12)$$

in the 2D case depicted in Figure 3, $\mathbf{J} = (0, 0, J_z)$. Then, $\mathbf{K}$ is imposed into the $A$ formulation as an external surface current density by means of boundary conditions of the form,

$$\hat{n} \times (\mathbf{H}_1 - \mathbf{H}_2) = \mathbf{K}, \qquad (13)$$

where $\hat{n}$ is the unit vector normal to the tape, and $\mathbf{H}_1$ and $\mathbf{H}_2$ are the magnetic field strength vectors above and below the HTS layer, respectively.

As in the case of the $H$ formulation, the selection of the elements used in the FEM matters. Two kind of elements are required, Lagrange second-order elements are used to approximate $\mathbf{A}$ and Lagrange first-order elements for $\mathbf{T}$, as justified in [47].

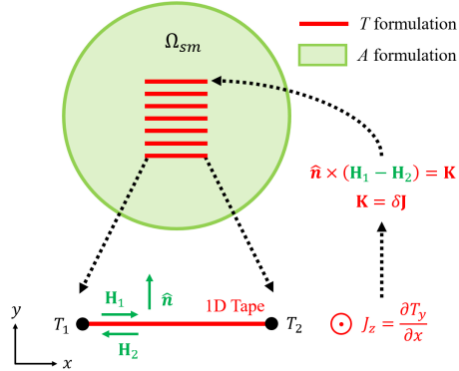

**Figure 3.** Bounded universe of the *T-A* formulation consisting of superconducting 1D layers and a surrounding medium. *T* is computed over the superconducting layers and *A* is computed over the entire bounded universe.

## 3. Case study and full models

### 3.1. Case study

The case study used in this work is the racetrack coil also used in [12], [47]. This coil has 10 pancakes, each consisting of 200 turns, bringing a total number of turns equal to 2000. The geometric parameters of the coil are summarized in Table 1. The symmetry of the coil allows modeling just one quarter of the coil's cross-section. Therefore, it is possible to consider only 5 stacks, each consisting of 100 turns in a planar 2D geometry. The coil, its cross-section and the modeled section are depicted in Figure 4.

The constitutive $E - J$ relation of the HTS material is modeled by the so-called power-law [50], therefore the resistivity of the superconducting subdomains $\Omega_{sc}$ is given by,

$$\rho_{HTS} = \frac{E_c}{J_c}\left|\frac{\mathbf{J}}{J_c}\right|^{n-1}. \tag{14}$$

The so-called Kim-like model [13], [51] is used to describe the anisotropic behavior of the HTS tapes, therefore $J_c$ is defined by,

$$J_c(\mathbf{B}) = \frac{J_{c0}}{\left(1 + \frac{\sqrt{k^2 B_\parallel^2 + B_\perp^2}}{B_0}\right)^\alpha}, \tag{15}$$

where $B_\perp$ and $B_\parallel$ are the magnetic flux density components perpendicular and parallel to the wide surface of the tape, respectively. The parameters of (14) and (15) are summarized in Table 2.

**Table 1.** Case study geometric parameters.

| Parameter | Value |
|---|---|
| Number of pancakes | 10 |
| Turns per pancake | 200 |
| Unit cell width | 4.45 mm |
| Unit cell thickness | 293 µm |
| HTS layer width | 4 mm |
| HTS layer thickness $\delta$ | 1 µm |

**Table 2.** Case study electromagnetic parameters.

| Parameter | Value |
|---|---|
| $E_c$ | $1 \times 10^{-4}$ Vm$^{-1}$ |
| $n$ | 38 |
| $J_{c0}$ | $2.8 \times 10^{10}$ Am$^{-2}$ |
| $B_0$ | 0.04265 T |
| $k$ | 0.29515 |
| $\alpha$ | 0.7 |

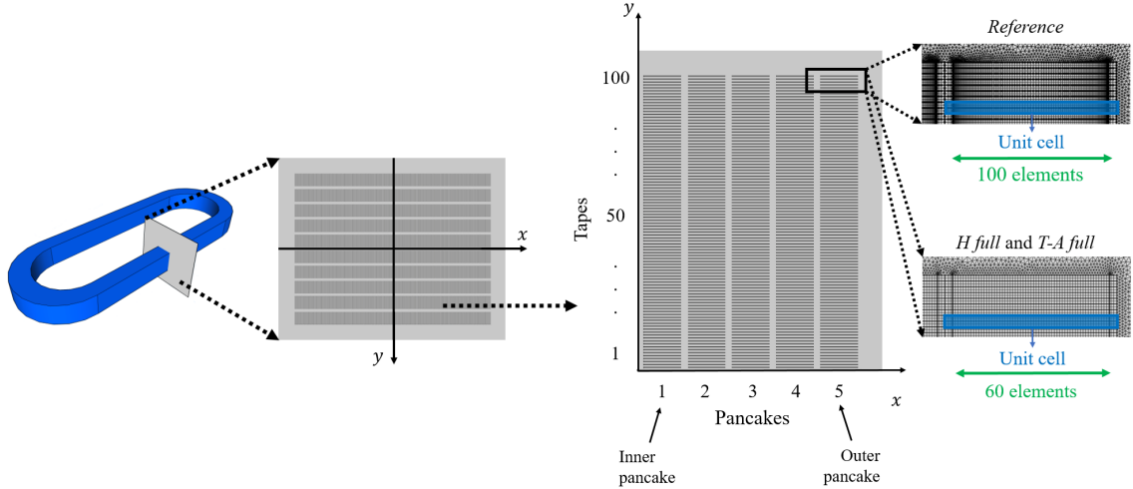

**Figure 4.** The racetrack coil used as case study has 10 pancakes with 200 turns per pancake. The coil can be modeled considering only ¼th of the coil's cross-section. The mesh in the unit cells is structured. The reference model considers 100 elements along the tape's width, while the *H* full and *T-A* full models consider 60 elements.

## 3.2. Reference model

The *H* formulation model that considers in detail each system's tape, presented in [12], [47] is used in this article to validate the rest of the models, and it is hereinafter called *reference model*.

The HTS tapes are composed by one layer of superconductor and different layers of normal conductors e.g., copper, silver, substrate [1]. The resistivity of the superconducting layer is several orders of magnitude lower than the resistivity of the other normal conductor layers forming part of the HTS tapes [14]. Therefore, the reference model does not include any normal conductor subdomain $\Omega_n$. The resistivity of the surrounding medium subdomain $\Omega_{sm}$ is considered to be $\rho_{sm} = 1$ $\Omega$m, [14]. No magnetic materials are considered, then the permeability of the superconducting $\Omega_{sc}$ and surrounding medium $\Omega_{sm}$ subdomains is equal to the permeability of vacuum $\mu_0$.

Figure 4 depicts the geometry of the reference model including the numbering of pancakes and tapes. The mesh of the *unit cells* is structured, and considers one element along the HTS layer's thickness and 100 elements along their width. An increasing number of elements towards the edges of the tape is considered to allows the accurate estimation of the **J** distribution in the regions where the magnetic field penetrates [12].

## 3.3. *H* and *T-A* full models

An assessment of the number of elements along the tape´s width was presented in [47]. The results demonstrate that, for the test conditions, the compromise between accuracy and computation time is fulfilled with 60 elements. The *H* formulation model that considers in detail each tape of the system and uses 60 elements along the tapes' width is hereinafter called *H full model*. The mesh of the unit cells of the *H* full model is also presented in Figure 4. Accordingly, the difference between the reference and the *H* full models is the distribution and the number of elements along the tapes' width, 100 elements with a non-uniform distribution for the reference model, and 60 elements with a uniform distribution for the *H* full model. In addition, throughout the rest of this work, it is assumed that all models have 60 elements along the tapes' width, with the exception of the reference model.

The *T-A full model* uses the *T-A* formulation, and as well as the reference and *H* full models, considers in detail all the tapes. Specifically, in the case of the *T-A* formulation models, "considers in detail all the tapes" means that the current vector potential **T** is computed along every single-tape. The mesh of the unit cells is also structured as shown in Figure 4. In this case, the HTS layers have no thickness and the mesh is made of 1D elements uniformly distributed along the HTS layers' width.

### 3.4. Results

The reference, *H* full and *T-A* full models were simulated considering one cycle of a sinusoidal transport current with an amplitude of 11 A, and a frequency of 50 Hz. The value 11 A was chosen because at the peak of the cycle the tape 1 of pancake 5 is completely penetrated by the current density. The simulation results are compiled in Figure 5, in tabular format. The first column contains the normalized current density $\mathbf{J}_n = \mathbf{J}/J_c$. The magnitude of the magnetic flux density $|\mathbf{B}|$ is presented in the second column. The plots of these first two columns show the results at the second peak of the transport current $t = 15$ ms. It is to see that the $\mathbf{J}_n$ and $|\mathbf{B}|$ plots are, respectively, indistinguishable to the naked eye.

The third column of Figure 5 contains the average losses plots. The x-axis in these plots represents the tape's number. There are five curves in each plot, one for each pancake. The numbering of the tapes and pancakes follows the order presented in Figure 4. The losses estimated with the *H* and *T-A* full models are very similar to those estimated with the reference model, but there are visible differences, particularly in the first two pancakes. Due to the higher current penetration, the losses in pancake 5 are almost three orders of magnitude larger than the losses in pancake 1. Although there are variations in the losses at the end of the pancakes, the losses in a given pancake remain within the same order of magnitude.

The quantitative comparison of the models is carried out by calculating the relative error of the average losses, the coefficient of determination $R^2$ of the **J** distributions, and the normalized computation time. These data are compiled in the fourth column of Figure 5.

The average losses are obtained using data of the second half of the cycle, as follows,

$$Q_{av} = \frac{2}{P} \int_{P/2}^{P} \iint_{\Omega_{sc}} \mathbf{E} \cdot \mathbf{J} \, ds \, dt, \tag{16}$$

where $P$ is the period of the sinusoidal cycle, and $\Omega_{sc}$ are the superconducting subdomains. The relative error of the average losses, expressed in percent, is defined as,

$$er_Q = \frac{(Q_{M\_av} - Q_{R\_av})}{Q_{R\_av}} \; 100 \; \%, \tag{17}$$

where $Q_{R\_av}$ and $Q_{M\_av}$ are the average losses computed with the reference model and the model that is being compared, respectively. For the test conditions, $Q_{R\_av} = 127.24$ W/m.

Unlike the average losses that are scalars, the **J** distributions are multivariable functions. The coefficient of determination is a widely used metric to evaluate the goodness of the fit [52], here it is used to compare the **J** distributions, and is defined as,

$$R^2 = 1 - \frac{\int_0^P \iint_{\Omega_{sc}} (\mathbf{J}_R - \mathbf{J}_M)^2 ds \, dt}{\int_0^P \iint_{\Omega_{sc}} (\mathbf{J}_R - \overline{\mathbf{J}}_R)^2 ds \, dt}, \tag{18}$$

where $\mathbf{J}_R$ and $\mathbf{J}_M$ are the **J** distributions computed with the reference and tested models, respectively. $\overline{\mathbf{J}}_R$ is the mean value of $\mathbf{J}_R$. It must be remembered that $R^2 = 1$, means a perfect matching between $\mathbf{J}_R$ and $\mathbf{J}_M$. The coefficient $R^2$ has an advantage over the error $er_Q$. The averaging nature of $er_Q$ tends to hide local and instantaneous errors, e.g., an instantaneous excess in the losses may be compensated by another instantaneous deficit; while these same errors have a cumulative effect in $R^2$.

The normalized computation time is defined as,

$$\overline{ct} = \frac{ct_M}{ct_R} \; 100 \; \%, \tag{19}$$

where $ct_R$ and $ct_M$ are the computation times required by the reference model and the model that is being compared, respectively. For the test conditions, $ct_R = 31$ h $32$ min.

The results of the fourth column in Figure 5 show that the accuracy of the *H* and *T-A* full models is satisfactory. The errors $er_Q$ are lower than 1.7 %, and coefficients $R^2$ are larger than 0.98. The computation time are 17 h 36 min ($\overline{ct} = 75.8$ %) and 3 h 14 min ($\overline{ct} = 10.25$ %), for the *H* and *T-A* full models, respectively. These values, more specifically the normalized computation times, demonstrate that the *T-A* formulation allows building more efficient full models.

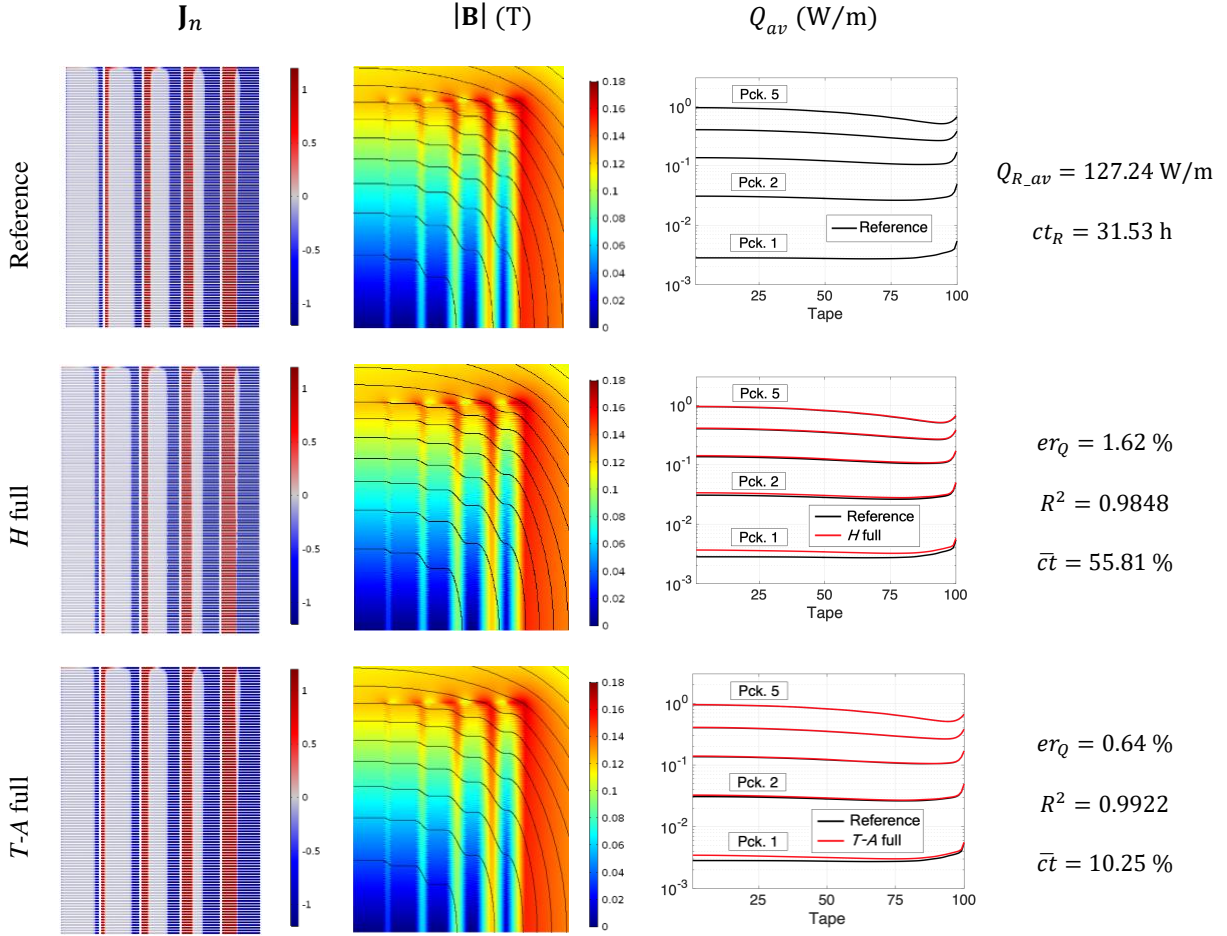

**Figure 5**. Results of the reference, *H* full, and *T-A* full models, in different rows respectively. The first and second columns show $\mathbf{J}_n$ and $|\mathbf{B}|$ at the second peak of the transport current ($t = 15$ ms). The third column shows the average losses as a function of the tape's number inside each pancake, the tapes and pancakes are numbered as indicated in Figure 4.

## 4. *H* formulation strategies

### 4.1. Homogenization

The homogenization method considers that the stacks of HTS tapes can be modeled as homogeneous bulks. As shown in Figure 6, a unit cell includes one HTS tape and its surrounding medium, e.g., cryogenic liquid and/or insulating materials. When the stack is transformed into an anisotropic bulk its geometric features are "washed out". This process is depicted in Figure 6. The model should include additional features that allow the electromagnetic behavior of the homogeneous bulk to resemble that of the original stack.

The non-superconducting materials forming part of the stack have resistivity values several orders of magnitude larger than those of the HTS material, hence only the HTS material resistivity is considered in the homogenization process. The resistivity of the bulk is derived from (14) and (15). But the $J_c$ value must be replaced by an engineering critical current density $J_{ce}$, defined as,

$$J_{ce} = f_{HTS} J_c, \qquad (20)$$

where

$$f_{HTS} = \frac{\delta}{\Delta}, \qquad (21)$$

$\delta$ is the real thickness of the HTS layer, and $\Delta$ is the thickness of the unit cell. Thus, the superconducting properties of the HTS tapes are diluted in the cross-section of the bulk.

In the $H$ full model, it is necessary one integral constraint per tape to impose the desired transport currents. In the $H$ homogenous model, the bulk subdomains $\Omega_h$ are further subdivided into bulk's subsets $\Omega_{h\_i}$, as depicted in Figure 6. Then, one constraint is necessary for each bulk's subset, instead of one constraint per tape. The losses are computed in each bulk's subset. The losses in each subset are divided by the number of tapes included in each subset, then it is possible to approximate the losses along the stack. A detailed description of the $H$ homogenous strategy can be consulted in [14].

The *H homogenous model* of the case study considers 5 bulks, one for each pancake. Each bulk is subdivided into 6 subsets, as shown in Figure 7. The subsets in the upper part of the pancakes have a larger aspect ratio than the ones closer to the symmetry plane. Such kind of distributions has been recommended in [12], [14]. The mesh of the bulks is structured considering one element along the subset's thickness and 60 elements along the tapes' width, as depicted in Figure 7.

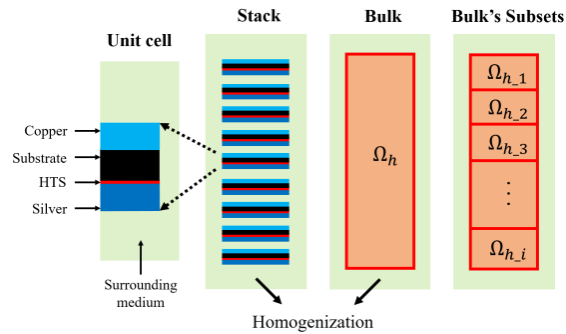

**Figure 6.** The homogenization process transforms a stack of HTS tapes into a homogeneous bulk. The bulk is subdivided into bulk's subsets $\Omega_{h\_i}$, one integral constraint is used to impose the transport current in each subset.

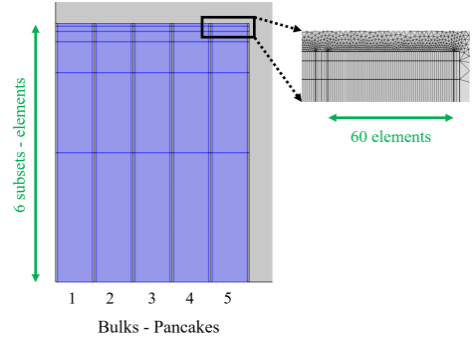

**Figure 7.** Geometry and mesh of the *H* and *T-A* homogeneous model. There are 5 bulks. Each bulk is subdivided into 6 subsets. The mesh of the bulks considers 1 element along the subsets' thickness and 60 elements along the tapes' width.

## 4.2. Densification

The *densification* method addresses the analysis of stacks by means of a reduced number of tapes, called *densified tapes*. The densified tapes merge a given number of tapes into a single tape. The densification can be understood as the opposite process to the homogenization. Instead of washing out the geometrical layout of the stack and dilute the transport current of the tapes in the bulk, here the densified tapes preserve their original geometry and concentrates the transport current of the merged tapes.

As in the previous models, the densified model does not include the normal conductors forming part of the HTS tapes. The resistivity of superconducting subdomains of the densified tapes is derived from (14) and (15). But, the $J_c$ value is replaced by a densified critical current density $J_{cd}$, defined as,

$$J_{cd} = dJ_c, \qquad (22)$$

where $d$ is the number of tapes merged into a single densified tape.

The transport current in the densified tapes is $I_{tr} = dI_k$, were $I_k$ is the transport current in the original non-densified tapes. It is necessary one integral constraint per densified tape. Then, as in the case of the *H* homogenous model the number of constraints is reduced.

The densification process is depicted in Figure 8. In this example, a given densified tape is built of 3 tapes, labelled $r - 1$, $r$ and $r + 1$, therefore $d = 3$. The densified tape is located at the position of the original tape $r$. It is not necessary for $d$ to be an integer number. The parameter $d$ may be equal to other real positive number. For instance, a stack made of 5 tapes can be modeled by means of 2 densified tapes. In this case, the densified tapes may merge 3 and 2 tapes, respectively. In another possible scenario, the densified tapes may merge 2.5 tapes each one, then the parameter $d$ should be $d = 2.5$ for both densified tapes.

Once the **J** distribution is computed, the losses can be calculated in the densified tapes. The losses in the densified tapes are divided by their corresponding $d$, and these values are used to approximate by interpolation the losses in each tape of the original stack.

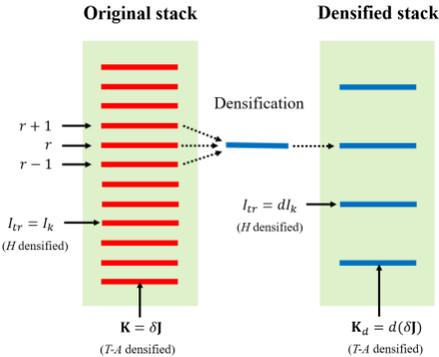

**Figure 8.** Densification process, the number of tapes in the original stack is reduced and the transport current of the merged tapes is forced to flow in the densified tapes.

In order to build a successful $H$ densified model of the case study, different sets of densified tapes were tried. According to our criterion, the compromise between accuracy and computation time is fulfilled with a set of 31 densified tapes per pancake. The geometry of the model and the position of the densified tapes is depicted in Figure 9. The first twenty one densified tapes merge four tapes each ($d = 4$). For the following five tapes, the parameter $d$ takes the values $\{3, 3, 2, 2\}$, respectively. Finally, the upper six densified tapes merge just one tape, ($d = 1$). The denser distribution of densified tapes at the upper part of the pancake allows achieving the required accuracy in the regions with larger variations in the **J** distributions.

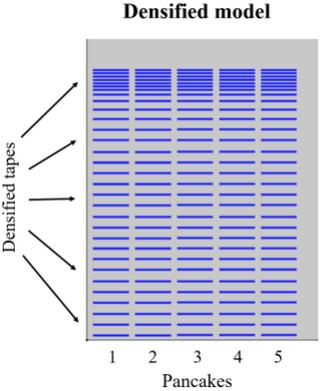

**Figure 9.** Geometry of the $H$ and $T$-$A$ densified models. There are 31 densified tapes in each pancake. A larger number of densified tapes is considered in the upper part of the pancakes.

## 4.3. Multi-scaling

The idea of the multi-scaling method is to break up the model into several smaller models. In this way, it is possible to reduce the size of the problem by analyzing in detail a subset of significant tapes called *analyzed tapes*.

The multi-scale models, as describe in [12], are formed by two 2D submodels. The first submodel is an *A* formulation magnetostatic model of the full coil including all the tapes with their actual geometry. This submodel, called *coil submodel*, does not consider any superconducting properties, hence the results depend on a predefined $\mathbf{J}_a$ distribution. The second submodel, called *single-tape submodel*, is an *H* formulation model of a unit cell containing just one tape. The single-tape submodel does not consider the normal conductor layers of the HTS tape, and the HTS layer is considered with its actual thickness. Both submodels are depicted in Figure 10.

The computational process is carried out in two steps. The first step is to use the coil submodel to estimate the background magnetic field strength **H** all across the bounded universe. Then, the **H** field along the boundary of the unit cells of the analyzed tapes is exported to the single-tape submodel as a time-dependent Dirichlet boundary conditions. The second step of the computational process is the use of the single-tape submodel. In this second step, the losses in all the analyzed tapes are calculated. Finally, the losses in the non-analyzed tapes are obtained by interpolation.

Breaking up the model into several smaller models not only reduces the computational burden, but also allows the parallelization of the problem, further reducing the computation time. A detailed description of the *H* multi-scale strategy can be consulted in [12].

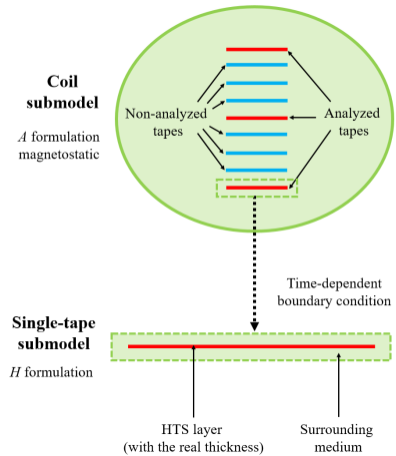

**Figure 10.** Coil and single-tape submodels. Both models consider the actual dimensions of the HTS layers (width and thickness).

The *H multi-scale model* of the case study uses 6 analyzed tapes in each pancake, 30 analyzed tapes in total. The distribution of the analyzed tapes is selected to be analog to the distribution bulk's subsets in the *H* homogenous model. This means that the positions of the analyzed tapes correspond with the center of each bulk's subset. The set of analyzed tapes in each pancake is {25, 66, 88, 96, 99, 100}. The position of the analyzed tapes is shown in Figure 11 (a). The distribution of analyzed tapes also respect the directives proposed in [12], [46].The mesh of the unit cells is structured and considers one element along the HTS layer's thickness and 60 elements along their width.

## 4.4. Iterative multi-scaling

The accuracy of the multi-scale models depends on the accuracy on the background magnetic field, which in turn depend on the predefined $\mathbf{J}_a$ distribution. The lack of knowledge of the predefined $\mathbf{J}_a$ distribution is the main limitation of the $H$ multi-scale strategy [12]. To address this issue and preserve the capability to analyze the system by means of a reduced set of analyzed tapes, the iterative multi-scale strategy was proposed in [46]. The iterative multi-scale strategy is the iterative implementation of the multi-scale strategy. The iterative multi-scale strategy allows obtaining a new and more accurate dynamic solution at each iteration.

In the multi-scale models, the some part of the information (background magnetic field) goes from the coil submodel to the single-tape submodel. Whereas, in the iterative strategy, another part of the information (current density) is allowed to flow back from the single-tape submodel to the coil submodel. To initialize the procedure, the $\mathbf{J}$ distribution in every tape is supposed to be uniform, then the coil submodel is used to estimate the background magnetic field. The $\mathbf{H}$ field along the boundary of the analyzed tapes is exported as time-dependent boundary condition to the single-tape submodel. Now, the single-tape submodel is not only used to compute the losses but also the current density. An interpolation method is used to estimate the $\mathbf{J}$ distributions in the non-analyzed tapes. The new $\mathbf{J}$ distribution for all the tapes is exported to the coil submodel and a new background magnetic field is computed. The process is repeated to obtain more accurate estimations.

To exit from the iterative loop, the $\mathbf{J}$ distribution for all the tapes of the current iteration is compared with the distribution of the previous iteration. The error of the $\mathbf{J}$ distributions at the iteration $k$ is defined as,

$$eJ_k = \frac{\sqrt{\int_0^P \iint_{\Omega_{sc}} (\mathbf{J}_{k-1} - \mathbf{J}_k)^2 \, ds \, dt}}{\sqrt{\int_0^P \iint_{\Omega_{sc}} (\mathbf{J}_{k-1})^2 \, ds \, dt}} \qquad (23)$$

where $\mathbf{J}_{k-1}$ and $\mathbf{J}_k$ are the $\mathbf{J}$ distributions for the iteration $k$ and $k-1$, respectively. If the error $eJ_k$ is smaller than a user-predefined criterion $\varepsilon$, then the process is completed. A detailed description of the $H$ iterative multi-scale strategy can be consulted in [46].

The *H iterative multi-scale model* uses the same set of analyzed tapes used by the $H$ multi-scale model, therefore Figure 11 (a) also represents the geometry of coil submodel of the $H$ iterative multi-scale model. The iterative multi-scale strategy requires the interpolation of the $\mathbf{J}$ distributions, the linear and the Inverse Cumulative Density Function (ICDF) interpolation methods were used. The

ICDF interpolation method [53] was adapted to interpolate **J** distribution in [46], the method produces more realistic distributions, avoiding some issues produced by the conventional linear interpolation.

## 4.5. Simultaneous multi-scaling

As describe above, *H* multi-scale and the *H* iterative multi-scale strategies use two different submodels. The computation of the **J** distributions in the analyzed tapes using the single-tape submodel can only be performed after the computation of the background magnetic field using the coil submodel. Therefore, the computation of the **J** distributions and the background field is not carried out simultaneously.

In this section, a new strategy called *simultaneous multi-scale* is proposed. The simultaneous multi-scale strategy allows simultaneously solving the **J** distribution and the background magnetic field. The strategy relies on the possibility to include an additional contribution in the Ampère's law (1). This summand allows imposing an external current density $\mathbf{J}_e$ in the superconducting subdomains $\Omega_{sc}$ of the non-analyzed tapes, as follows,

$$\nabla \times \mathbf{H} = \mathbf{J} + \mathbf{J}_e. \tag{24}$$

Considering Faraday's law (2) and the constitutive relations of the materials, the governing equation of the *H* formulation is then expressed as,

$$\nabla \times (\rho(\nabla \times \mathbf{H} - \mathbf{J}_e)) = -\mu \frac{\partial \mathbf{H}}{\partial t}. \tag{25}$$

The external $\mathbf{J}_e$ in the superconducting subdomains $\Omega_{sc}$ of the analyzed tapes and in the surrounding medium subdomain $\Omega_{sm}$ is zero. The external $\mathbf{J}_e$ in the superconducting subdomains $\Omega_{sc}$ of the non-analyzed tapes is approximated by interpolating the **J** distributions of the analyzed tapes.

The resistivity in the superconducting subdomains $\Omega_{sc}$ of the analyzed tapes is defined by (14) and (15). The resistivity of the superconducting subdomains $\Omega_{sc}$ of the non-analyzed tapes is considered to be the resistivity of the surrounding medium, $\rho_{sm} = 1$ Ωm. This value is orders of magnitude larger than the resistivity of the superconducting subdomains, therefore the induced current density in the non-analyzed tapes has a negligible impact when compared with the external $\mathbf{J}_e$.

The *H simultaneous multi-scale model* of the case study considers the same set of 30 analyzed tapes of the *H* multi-scale model. The non-analyzed tapes in the *H* simultaneous multi-scale model preserve their original geometry. Hence, Figure 11 (a) also represents the geometry of the *H* simultaneous multi-scale model. It is possible to reduce the number of DOF and the computational burden of the *H* simultaneous multi-scale model by means of the homogenization or densification of the non-analyzed tapes. Therefore, two additional models are presented here: the *H simultaneous multi-scale homogenous model* and the *H simultaneous multi-scale densified model*.

In the $H$ simultaneous multi-scale homogenous and $H$ simultaneous multi-scale densified models not all the non-analyzed tapes are homogenized or densified. The non-analyzed tapes adjacent to the analyzed tapes keep their original shape. These non-homogenous/non-analyzed or non-densified/non-analyzed tapes are used to establish a greater distance between the analyzed tapes and the distortions in the magnetic field produced by the homogeneous or densified tapes. The geometries of the three $H$ simultaneous multi-scale models are presented in Figure 11.

The implementation ICDF interpolation, as presented in [46], requires the computation of integrals, derivatives and inverse functions. In the $H$ iterative multi-scale model, this method is implemented in a MATLAB script. The $H$ simultaneous multi-scale models were implemented in a single COMSOL model, and for convenience just the simpler linear interpolation was used. This is not a major drawback because the ICDF interpolation makes only a marginal contribution to the accuracy of the model [46].

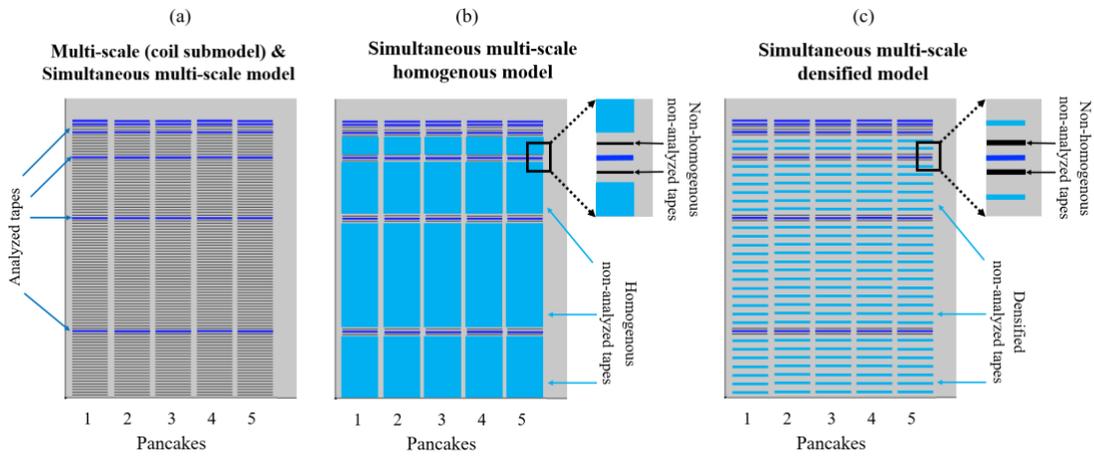

**Figure 11.** (a) geometry of the coil submodel of the multi-scale and iterative multi-scale model, and the simultaneous multi-scale models. There are 6 analyzed tapes per pancake. (b) Geometry of the simultaneous multi-scale homogenous model, most of the non-analyzed tapes are homogenized. (c) Geometry of the simultaneous multi-scale densified model, most of the non-analyzed tapes are densified.

### 4.6. Results

To compare and validate the strategies described in this section, the models were also simulated considering a sinusoidal transport current (11 A, 50 Hz). The results of the $H$ homogenous, $H$ densified, $H$ multi-scale and $H$ iterative multi-scale models are presented in Figure 12. The results of the three $H$ simultaneous multi-scale models are presented in Figure 13. These figures have the format of Figure 5. The first two columns show $\mathbf{J}_n$ and $|\mathbf{B}|$, both at $t = 15$ ms, the average losses are displayed in the third column, and the last column contains the quantitative data.

The first row in Figure 12 shows the results of the $H$ homogenous model. This model successfully reproduces the screening currents of the reference model. The $\mathbf{J}_n$ plot in this row clearly shows the homogenous regions inside the pancakes. The $|\mathbf{B}|$ distribution presents a smoother profile due to the homogenous current densities. The accurate estimation provided by this model is confirmed by the following values: $er_Q = 1.28\ \%$ and $R^2 = 0.9221$. In this case the coefficient $R^2$ is computed rescaling the current density (dividing by $f_{HTS}$) and considering just the values at the positions of the original superconductor subdomains. The computation time required by the $H$ homogenous model is 36 min 44 s ($\overline{ct} = 1.94\ \%$).

The second row in Figure 12 presets the results of the $H$ densified model. Thicker lines are used to represent the densified tapes. In contrast to the homogenous case, here the $|\mathbf{B}|$ distribution has a rougher profile. The accuracy is degraded due to the nature of the densified tapes, larger self-fields and larger distances between tapes. This degradation is reflected in the values $er_Q = 6.67\ \%$ and $R^2 = 0.8549$. For the purpose of computing $R^2$, the $\mathbf{J}$ distributions of the densified tapes are rescaled (divide by $d$) and the $\mathbf{J}$ distributions in the removed tapes are approximated using linear interpolation. The computation time is reduced compared to the $H$ full model, but this reduction is not as big as in the case of the $H$ homogenous model, the normalized computation time is $\overline{ct} = 29.03\ \%$.

The results of the $H$ multi-scale model are presented in the third row of Figure 12. The predefined current density distribution $\mathbf{J}_a$ is uniform, as can be seen in the first entry of the third row. The uniform $\mathbf{J}_a$ does not contain any screening current, then it is not the best approximation of the reference $\mathbf{J}$ distribution. This fact is also reflected in the low coefficient $R^2 = 0.0304$. Consequently, the magnetic flux density and the losses exhibit noticeable errors, especially at in the upper part of the pancakes. The red circles in the plot of the losses indicate the position of the analyzed tapes, this is also the case of all the multi-scale models. The losses error is $er_Q = -21.7\ \%$, the negative sign indicates that the losses are underestimated. The computation time is 27 min 30 s ($\overline{ct} = 1.45\ \%$), this time is the summation of the time required to run the coil submodel one time (5 min) and the single-tape submodel 30 times, one for each analyzed tape (the average computation time of the single-tape submodel is 45 s).

The results of the $H$ iterative multi-scale model are compiled in the last row of Figure 12. The convergence criterion is defined as $\varepsilon = 0.01$, this criterion is reached at the $7_{th}$ iteration regardless of the interpolation method (linear of ICDF) used to approximate the $\mathbf{J}$ distributions in the non-analyzed tapes. Figure 12 presents the results when the ICDF interpolation is used. The results when the linear interpolation is applied are visually indistinguishable to those obtained from the ICDF interpolation, then for clarity only the latter are shown. The error $er_Q$ when linear interpolation is used is $-0.87\ \%$, when ICDF interpolation is used $er_Q = -0.56\ \%$. The coefficient $R^2$ takes values 0.9796 and 0.9803, for the linear and ICDF interpolations, respectively. The accuracy is marginally better with the ICDF interpolation. The computation time is 3 h 21 min ($\overline{ct} = 10.51\ \%$) with linear interpolation, and 3 h 17 min ($\overline{ct} = 10.33\ \%$) with ICDF interpolation. These times are approximately seven times the computation time of the $H$ multi-scale model.

The rows of Figure 13 compiles the results of the $H$ simultaneous multi-scale, $H$ simultaneous multi-scale homogenous, and $H$ simultaneous multi-scale densified models, respectively. The plots in the first column allow the observation of the different approaches in which the non-analyzed tapes are modeled: tapes with its original geometry, homogenous bulks or densified tapes. The $|\mathbf{B}|$ distribution of the $H$ simultaneous multi-scale is visually indistinguishable to the reference $|\mathbf{B}|$ distribution. Moreover, it is easy to find similarities between the distortions in the $|\mathbf{B}|$ distributions of the $H$ homogenous and $H$ simultaneous multi-scale homogenous models; and between the distortions of the $H$ densified and $H$ simultaneous multi-scale densified models. The three models have acceptable and similar accuracies, as demonstrated by the $er_Q$ values lower than 1.6 %, and the $R^2$ values greater than 0.98. The advantage of the simplified description of the non-analyzed tapes is clearly observed in the computation times. The computation times of the $H$ simultaneous multi-scale is 16 h 56 min ($\overline{ct} = 53.7$ %), this computation time is halved with the other two models.

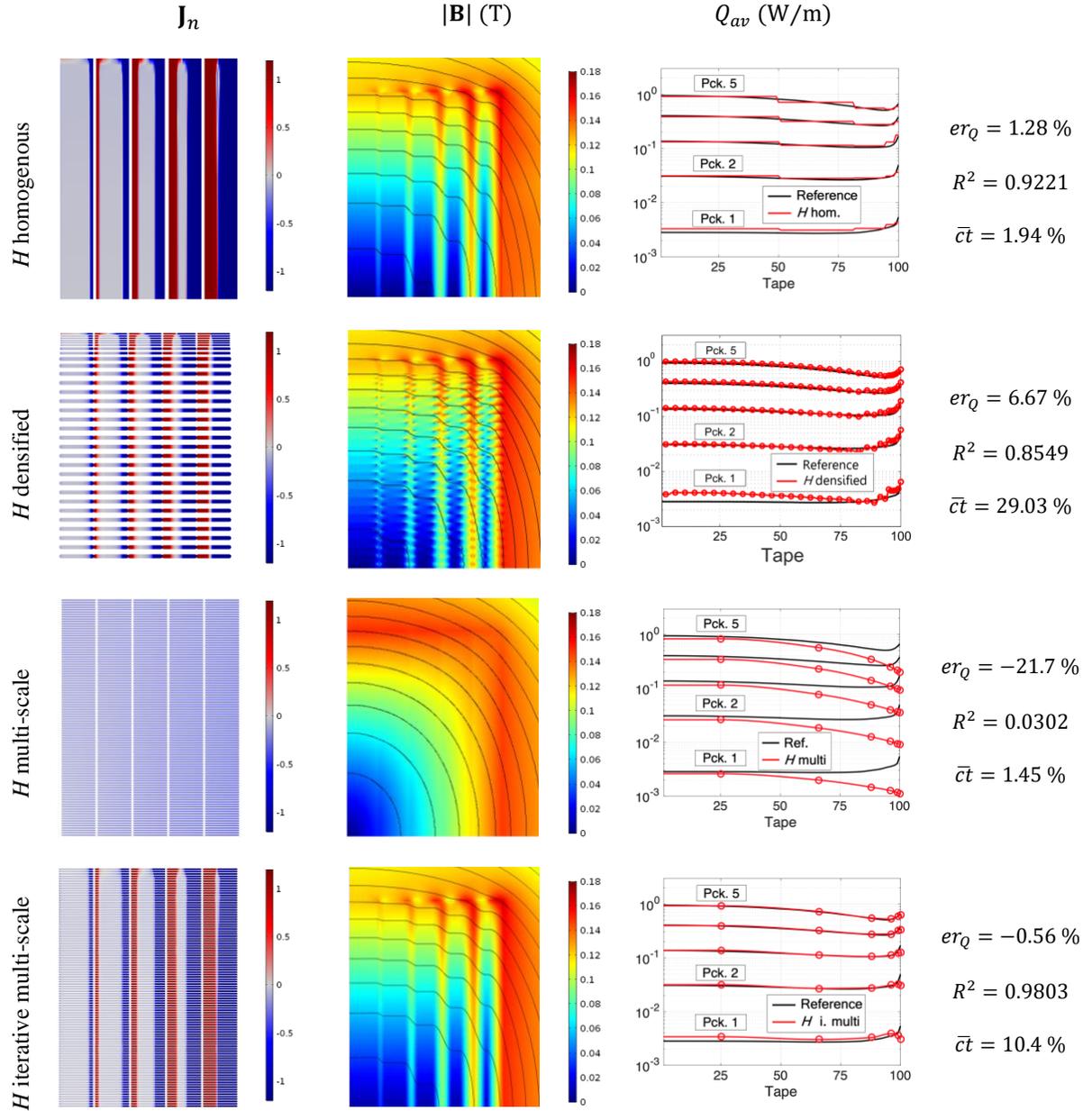

**Figure 12.** Results of $H$ homogeneous, $H$ densified, $H$ multi-scale and $H$ iterative multi-scale and models. The first and second columns show, respectively, $\mathbf{J}_n$ and $|\mathbf{B}|$ at the second peak of the transport current ($t = 15$ ms). The third column shows the average losses as a function of the tape's number inside each pancake.

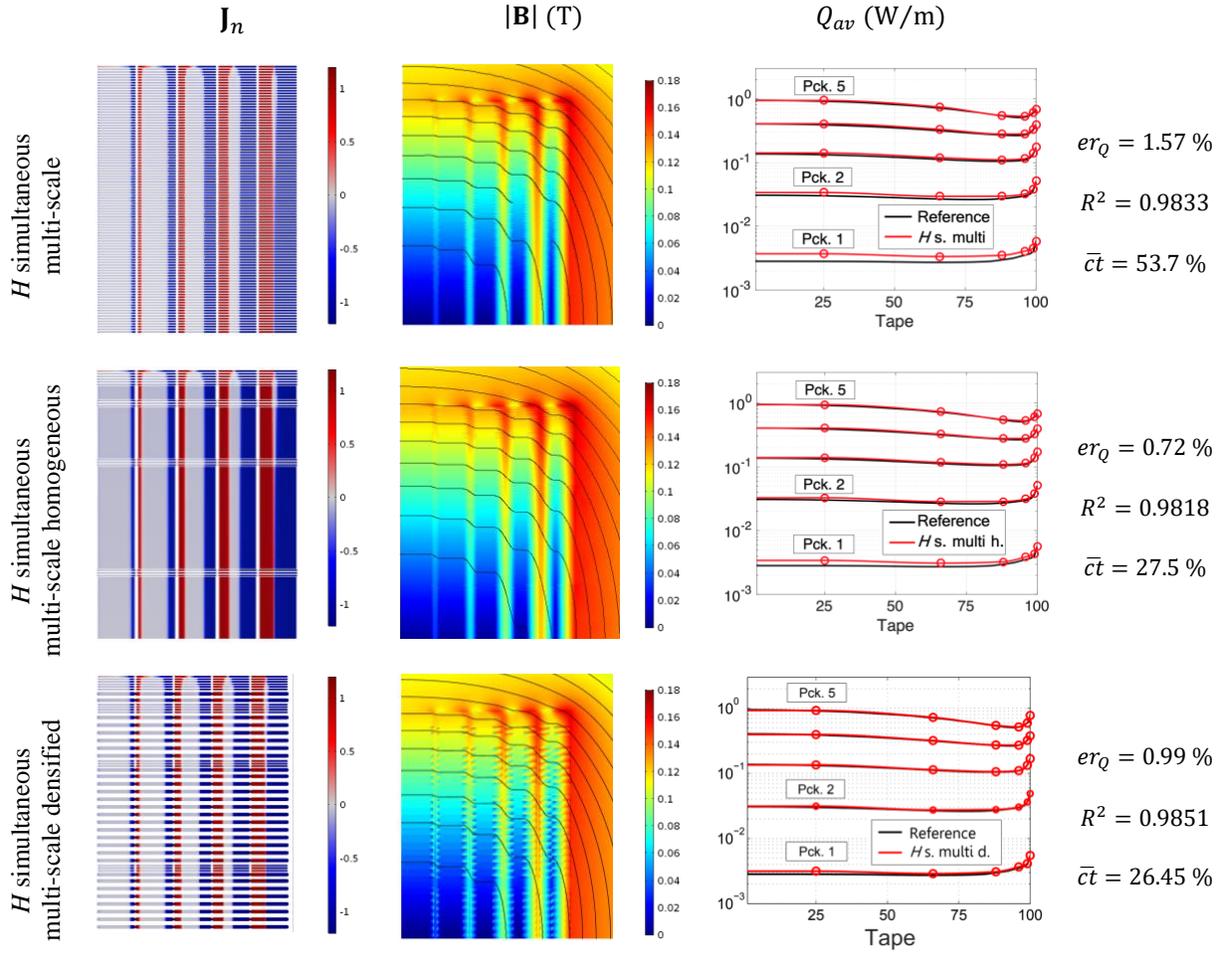

**Figure 13.** Results of the *H* simultaneous multi-scale, *H* simultaneous multi-scale homogenous, and *H* simultaneous multi-scale densified models. The first two columns present the results at the second peak of the transport current ($t = 15$ ms). The third column shows the average losses.

## 5. *T-A* formulation strategies

### 5.1. Homogenization

The manner in which the homogenization is coupled with the *T-A* formulation is depicted in Figure 14. The magnetic vector potential **A** is defined all over the entire bounded universe. The stacks of 1D HTS layers are transformed into 2D bulks, and the potential **T** is exclusively defined inside the bulk. For the purpose of computing **T**, the influence of the component of **B** parallel to the surface of the tapes ($B_x$ in the case of Figure 14) is not considered. Therefore, from equation (7), it follows that **T** has only one non-zero component ($T_y$ in the case of Figure 14) that is defined by means of equation (8).

The homogenous bulk can be understood as the limiting case of a densely packed stack made of 1D HTS layers. Each 1D layer of the densely packed stack (homogenous bulk) must carry the same amount of infinitesimal transport current. To impose such transport current, it is necessary to use the values $T_1$ and $T_2$, defined in (9), as Dirichlet boundary conditions along the edges of the bulks perpendicular to the tapes. To compensate for the fact that the current density is computed inside the homogenous bulk, a new engineering current density $J_{ze}$ is defined as,

$$J_{ze} = f_{HTS} J_z \tag{26}$$

where $f_{HTS}$ is the ratio defined in (21). $J_{ze}$ is imposed as a source into the *A* formulation, then equation (10) is transformed into,

$$\nabla^2 A_z = -\mu J_{ze}. \tag{27}$$

For the purpose of computing **T**, the resistivity of the bulk subdomains is considered to be the resistivity of the superconducting material, defined by (14) and (15). The losses are computed by integrating the local losses along the lines parallel to the HTS layers at the center of each bulk's subset, then the losses along the rest of the tapes are approximated by interpolation. A more detailed description of the *T-A* homogenous strategy can be consulted in [47].

The *T-A homogenous model* of the case study, as well as the *H* homogenous model, considers 5 bulks. Here, the bulk's subsets of the *H* homogenous model are not required to impose the transport current by means of integral constraints; but these subsets are used to define the distribution of the elements along the bulk's height. The mesh of the bulks in the *T-A* homogeneous model is structured and considers one element along the subsets thickness and 60 elements along the tapes' width. Figure 7 also represents the geometry of the *T-A* homogenous model.

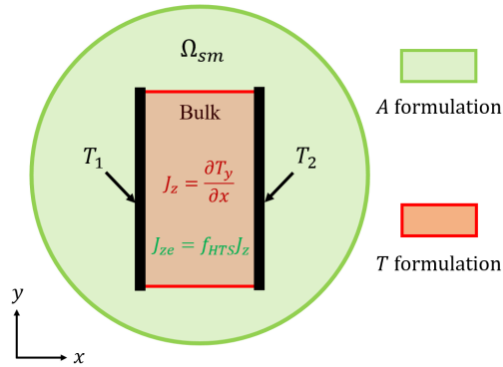

**Figure 14.** *T-A* homogeneous strategy. *A* is defined in the entire bounded universe, and *T* is exclusively defined inside the bulk. The boundary conditions $T_1$ and $T_2$ are applied to the vertical edges of the bulk. The engineering current density $J_{ze}$ is imposed into the *A* formulation.

## 5.2. Densification

The densification method can also be used in conjunction with the *T-A* formulation. Here also, the idea is to model the original stack by means of lower number of densified tapes.

In a *T-A* densified model the densified tapes are one-dimensional objects along which the **J** distributions are computed by means of the current density potential **T**, and its only non-zero component $T_y$, as given in equation (8). The resistivity o the HTS material is defined by means of (14) and (15). Unlike the *H* densified strategy, here the critical current density is not modified. In the *T-A* formulation, the surface current density $\mathbf{K} = \delta \mathbf{J}$ is imposed into the *A* formulation, see equations (12) and (13). To take into account the densification, the surface current density to be imposed is now defined as,

$$\mathbf{K}_d = d(\delta \mathbf{J}), \tag{28}$$

where $d$ is the number of tapes merged into a single densified tape.

The *T-A* densification process also depicted in Figure 8 involves two steps. The first step is to remove the tapes $r-1$ and $r+1$. Second, the magnetic effect of tape $r$ is forced to be three time larger than the magnetic effect of the same tape in the *T-A* full model. The transport current in the densified tape remains the same as the transport current in original tape, see equation (9). The magnetic effect of the densified tapes is incremented by means of the parameter $d$ in equation (28). The losses can be calculated in the densified tapes, these losses are considered as the losses produced by the tapes of the original stack in the position of the densified tapes. The losses in the removed tapes are approximated by interpolation.

The *T-A* densified model of the case study considers the set of densified tapes of the *H* densified model, 31 densified tapes per pancake. Then, Figure 9 also represents the geometry of the *T-A* densified model.

## 5.3. Simultaneous multi-scaling

The *T-A simultaneous multi-scale* strategy, as well as the *H simultaneous multi-scale* strategy, does not require two different submodels. The computation of the background magnetic field and the **J** distribution are carried out in a single model based on the *T-A* formulation.

In the *T-A* full models, the current vector potential **T** is defined over all the tapes. In the present approach, **T** is defined only along the analyzed tapes. The **J** distribution along the analyzed tape is obtained by calculating **T**, see equation (8). The **J** distributions in the non-analyzed tapes are approximated by linear interpolation using the **J** distributions of the analyzed tapes. The magnetic potential **A** is defined over the entire bounded universe. The current density in both analyzed and non-analyzed tapes is multiplied by the thickness of the superconducting layer $\delta$ to obtain a surface current density **K** to be imposed into the *A* formulation, see equations (11) and (13).

As it was the case with the *H* simultaneous multi-scale models, the DOF can be reduced by means of the homogenization or densification of the non-analyzed tapes. Therefore, three *T-A* simultaneous multi-scale models are presented. The difference between these models is the treatment of the non-analyzed tapes. The three models use the same set of 6 analyzed tapes per pancake used in the *H* multi-scale models. The first model, called *T-A simultaneous multi-scale*, considers the non-analyzed tapes with their original number and geometry.

As justified in [47], the *T-A* formulation uses first order elements to approximate **T**, and second order elements to approximate **A**. If first order elements are used for both quantities, the computation time can be reduced [45], but this choice produces undesired spurious oscillations in the **J** distributions [47]. To increase the computational efficiency without compromising the accuracy, the unit cells of the analyzed tapes and their adjacent non-analyzed tapes use second order elements to approximate **A**, while first order elements are used to approximate **A** throughout the rest of the system. Additionally, 30 elements are considered along most of the non-analyzed tapes, while 60 elements are considered in the analyzed tapes and their adjacent non-analyzed tapes. A more detailed description of the *T-A* simultaneous multi-scale strategy can be consulted in [47].

The other two *T-A* multi-scale models simplify the geometric description of the pancakes by means of the homogenization or densification of the non-analyzed tapes. These models are called *T-A simultaneous multi-scale homogenous* and *T-A simultaneous multi-scale densified* models, respectively. As it was done in Section 4.5, the non-analyzed tapes adjacent to the analyzed tapes keep their original geometry to establish greater distance between the analyzed tapes and the distortions in the magnetic field produced by the densified or homogenized non-analyzed tapes. The geometries of the *T-A* multi-scale models correspond with the geometries of the *H* multi-scale models, and are shown in Figure 11.

### 5.4. Results

The *T-A* models were validated considering the same operating conditions used in the previous sections. The results are compiled in Figure 15, using the tabular format of Figure 5.

The first row in Figure 15 shows the results of the *T-A* homogenous model. Due to the homogenization process, a smoother |**B**| distribution can be observed. In the losses plot of this row, losses in the pancake 1 deviate from the reference results. The losses in the first pancake are two orders of magnitude lower than those of the pancake 5, therefore the deviation does not affect the global result. The total losses are slightly overestimated, the error is $er_Q = 0.71$ %. The computation time required by the *T-A* homogenous model is 14 min 51 s ($\overline{ct} = 0.78$ %). Then, this is the fastest model of all the models presented in this article.

The results of the *T-A* densified model are presented in the second row of Figure 15. The densification process causes a rougher |**B**| distribution. The coefficient $R^2 = 0.8854$ demonstrate that, among the *T-A* models, the worst approximation of the current density is achieved by the *T-A* densified model. However, it remains an acceptable strategy for estimating the losses, as demonstrated by the value $er_Q = -2.62\ \%$. In contrast to the *H* densified model, here the losses are slightly underestimated. The computation time of the *T-A* densified model is 61 min ($\bar{ct} = 3.22\ \%$), approximately one third of the computation time of the *T-A* full model.

The last three rows contain the results of the *T-A* simultaneous multi-scale models. The $\mathbf{J}_n$ and |**B**| distributions calculated with the *T-A* simultaneous multi-scale model is indistinguishable from the reference distributions. The |**B**| distributions of the *T-A* simultaneous multi-scale homogenous and densified models show the respective distortions produced by the simplification of the non-analyzed tapes. Accurate estimations are achieved with the three *T-A* simultaneous multi-scale models. In the three cases, the magnitude of the error $er_Q$ is less than 1 %, and the coefficient $R^2$ is greater than 0.98. The computation time of the *T-A* full model ($\bar{ct} = 10.25\ \%$) is halved by the *T-A* simultaneous multi-scale model ($\bar{ct} = 5.06\ \%$). The densification of the non-analyzed tapes further reduces the computation time by 30 min ($\bar{ct} = 3.46\ \%$). Conversely, the homogenization of the non-analyzed tapes produces a noticeable increment of the computation time ($\bar{ct} = 18.41\ \%$).

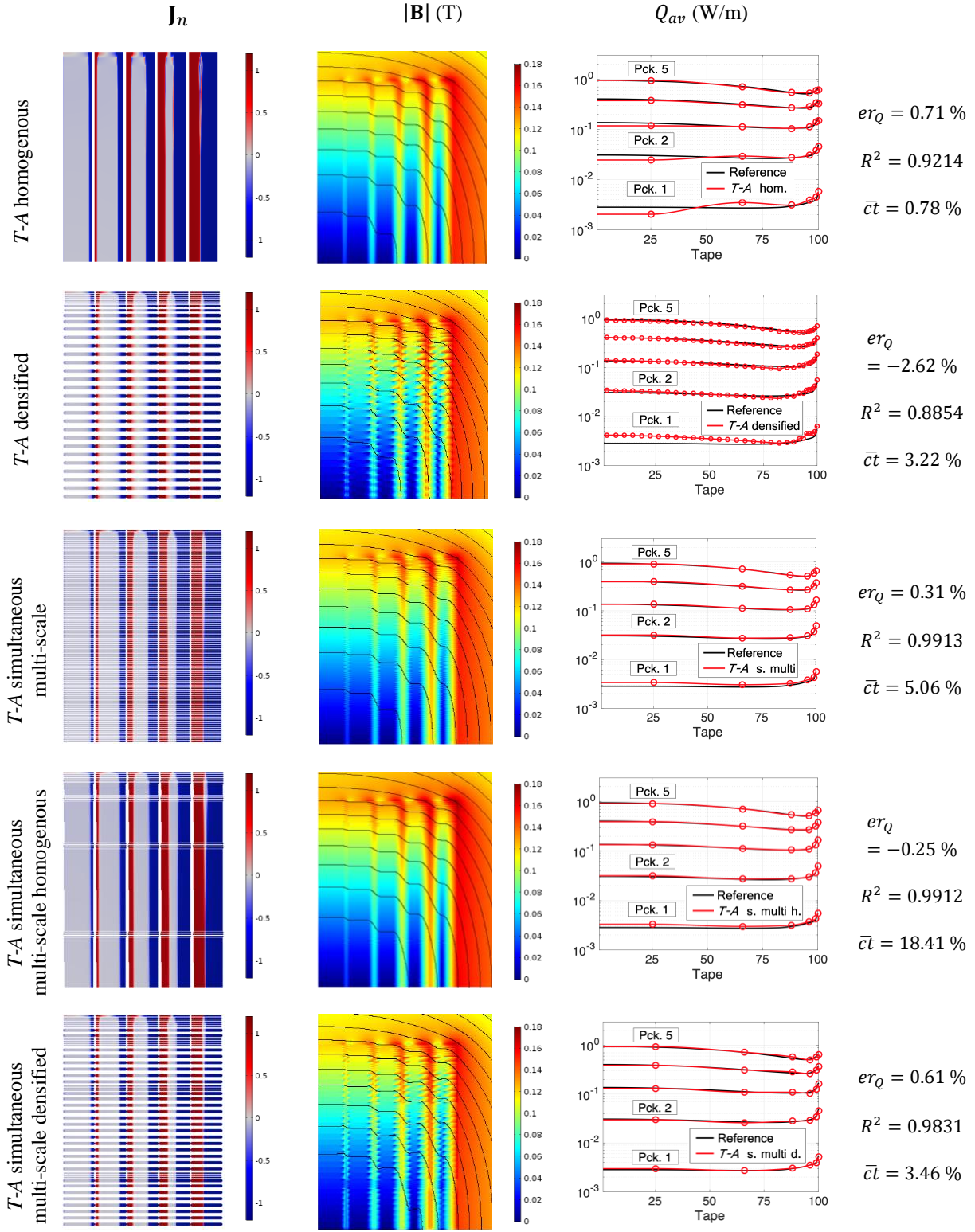

**Figure 15.** Results of the *T-A* model. The $J_n$ and $|B|$ plots show the results at the second peak of the transport current ($t = 15\ ms$). The third column shows the average losses. The three multi-scale models use the usual set of 6 analyzed tapes per pancake.

## 6. Comparison

For ease of comparison; the error $er_Q$, the coefficient $R^2$, and the normalized computation time $\overline{ct}$ are brought together in Table 3 to infer which is the most efficient strategy. Additionally, the number of DOF of each model are presented. The first comparison, arising from Table 3, is the comparison between the reference and the $H$ full models. The reference model is the model with the largest number of DOF and the largest computation time The reduction in the number of elements causes a reduction of the same order in the number of DOF. The speed up factor $(100/\overline{ct})$ due to the reduction of the number of elements is about 1.8.

The densification of the tapes in the $H$ densified model allows reducing the DOF and the computation time, but this model has the second lowest accuracy with a coefficient $R^2 = 0.8549$. The number of densified tapes in the densified model has to be larger than the number of analyzed tapes in the multi-scale models, otherwise the accuracy is drastically affected.

The model with the lowest number of DOF is the single-tape submodels of the $H$ multi-scale and $H$ iterative multi-scale models. The coil submodel has more DOF, and their computation is 5 min. But, the computation time of the $H$ multi-scale model is mostly determined by the size of the single-tape submodel and the number of analyzed tapes. Despite the repeated use of the single-tape submodel, the $H$ multi-scale model is the one with the lowest computation time among the models using the $H$ formulation. However, the accuracy of the $H$ multi-scale model is the lowest ($er_Q = -21.7$ %, $R^2 = 0.0304$). The accuracy is improved by the iterative implementation of the multi-scale strategy; but in exchange, the computation time increases. For the case study used here, the number of iterations is seven. Therefore, the computation time of the $H$ iterative multi-scale model is approximately seven times that of the $H$ multi-scale model.

The computation time of the $H$ simultaneous multi-scale model is similar to that of the $H$ full model; both models share the same variables, geometry, and number of DOF. The $H$ simultaneous multi-scale model prevents the presence of the non-linear resistivity of the superconductor material in the non-analyzed tapes, as described in Section 4.5. Nevertheless, this simplification in the non-analyzed tapes is not reflected in the computation time, possibly due to the imposition of the interpolated **J** distributions and the interpolation process itself. The homogenization and densification of the non-analyzed tapes in the $H$ simultaneous multi-scale homogeneous model reduces the number of DOF, with the larger reduction in the case of the homogenization. These reductions, even different, allows reducing the computation time roughly by a factor of two. The accuracy of the $H$ simultaneous multi-scale models is similar for the three cases, with coefficients $R^2$ larger than 0.98, and errors $er_Q$ lower than 1.6 %.

The $T$-$A$ full model has more DOF than the $H$ full model, this difference is mostly due to the use of second-order elements to approximate **A** in the $T$-$A$ formulation. Despite the larger number of DOF, the $T$-$A$ full model is approximately five times faster than the $H$ full model. Both full models consider the same number of elements along the tape's width, hence the reduction in the computation time is just influenced by the choice of the formulation.

Table 3. Models comparison.

| Model | Losses (W/m) | Comp. time (h) | DOF |
|---|---|---|---|
| Reference | 127.24 | 31 h 32 min | 563893 |

| Model | $er_Q$ (%) | $R^2$ | $\bar{ct}$ (%) | DOF |
|---|---|---|---|---|
| $H$ full | 1.62 | 0.9848 | 55.81 | 359408 |
| $H$ homogenous | 1.28 | 0.9221 | 1.94 | 11838 |
| $H$ densified | 6.67 | 0.8549 | 29.03 | 120424 |
| $H$ multi-scale | -21.7 | 0.0302 | 1.45 | cs-112752, st-709 * |
| $H$ iterative multi-scale | -0.56 | 0.9803 | 10.41 | cs-112752, st-709 * |
| $H$ simultaneous multi-scale | 1.57 | 0.9833 | 53.70 | 359408 |
| $H$ sim. multi-scale homogeneous | 0.72 | 0.9818 | 27.54 | 66752 |
| $H$ sim. multi-scale densified | 0.99 | 0.9851 | 26.45 | 158643 |
| $T$-$A$ full | 0.64 | 0.9922 | 10.25 | 548624 |
| $T$-$A$ homogenous | 0.71 | 0.9214 | 0.78 | 20612 |
| $T$-$A$ densified | -2.62 | 0.8854 | 3.22 | 103638 |
| $T$-$A$ simultaneous multi-scale | 0.31 | 0.9913 | 5.06 | 114582 |
| $T$-$A$ sim. multi-scale homogeneous | -0.25 | 0.9912 | 18.41 | 107529 |
| $T$-$A$ sim. multi-scale densified | 0.61 | 0.9831 | 3.46 | 112853 |

* The abbreviation cs stands for the DOF of the coil submodel, and the abbreviation st stands for single-tape submodel.

Among the $T$-$A$ models, the one with the lowest accuracy ($R^2 = 0.8854$) is the $T$-$A$ densified model. Comparing this coefficient with that of the $H$ densified model, the similarity indicates that there is a systematic degradation in the accuracy due to the densification process. In both cases, the accuracy reduction results from the distortions in the magnetic field produced by the separation between densified tapes.

When going from the $H$ full model to the $H$ simultaneous multi-scale model, the number of DOF remains the same; conversely when going from the $T$-$A$ full model to the $T$-$A$ simultaneous multi-scale model the number of DOF is reduced. This reduction is possible because in the $T$-$A$ simultaneous multi-scale model the vector potential **T** is not computed in the non-analyzed tapes, and the DOF associated with **T** in the non-analyzed tapes are not required. The superiority of the $T$-$A$ simultaneous multi-scale model over the $H$ simultaneous multi-scale becomes clear when the computation times are compared. The $T$-$A$ simultaneous multi-scale model is approximately ten times faster than the $H$ simultaneous multi-scale model. At the same time, the $T$-$A$ simultaneous multi-scale model is two times faster than $T$-$A$ full model. The accuracy of the $T$-$A$ simultaneous multi-scale model ($R^2 = 0.9913$) is slightly better than the $H$ simultaneous multi-scale model ($R^2 = 0.9833$).

The $T$-$A$ simultaneous multi-scale model can be simplified by means of either homogenous bulks or densified tapes to model the non-analyzed tapes. The accuracy of the three $T$-$A$ simultaneous multi-scale models is similar, with coefficients $R^2$ between 0.9831 and 0.9913. Indeed, the computation time is reduced in the *T-A* simultaneous multi-scale densified model, but in the *T-A* simultaneous multi-scale homogeneous model this time is more than three times the computation time of the *T-A* simultaneous multi-scale model.

Other than the single-tape submodels of the *H* multi-scale models, the models with the lowest number of DOF are the *H* and *T-A* homogenous models. When compared to their respective full models speed up factors induced by the homogenization are 28.8 and 13.1, for the *H* and *T-A* homogenous models, respectively. At the same time, the *T-A* homogenous model (15 min) is approximately 2.5 times faster than the *H* homogenous model (37 min). The *T-A* homogenous is the most efficient strategy, because it benefits from the speed up factors offered by the *T-A* formulation and the homogenization. Nevertheless, the homogenization causes some distortions in the **J** distribution; the presence such distortions is reflected in the reduction of the coefficients $R^2$ to around 0.92 for both homogeneous models.

When dealing with lager systems the speed up factors of the *T-A* homogenous models can be increased. For instance, the 32 T all-superconducting magnet from the NHMFL in Tallahassee has an HTS insert made of more than 20000 turns of REBCO tape [54]–[56], and there is not a published full model for such system. It is reported in [57], that the *T-A* homogenous model of the 32 T magnet [58] is 396.2 times faster than the *H* iterative multi-scale model of the same magnet [59].

## 7. Conclusions

In general, all the simplification strategies, i.e., homogenization, densification and multi-scaling, simplify the description of the systems and allow quicker results without seriously compromising the accuracy. The only case in which the computation time is not reduced is the *H* simultaneous multi-scale. And, the only case in which the accuracy is not satisfactory (compared with the rest of the models presented here) is the *H* multi-scale. Accordingly, the models using a simplification strategy represent a good alternative to the full models, an alternative that is more significant in the cases where the size of the system complicates the implementation and use of the full models.

As claimed in [43], the *H* formulation has a widespread popularity within the applied superconductivity community. However, the *T-A* formulation has proven to be more efficient to model systems made of HTS tapes. Accordingly, the strategies based on the *T-A* formulation have lower computation times than their *H* formulation counterparts. It is important to emphasize that the *T-A* homogenous model has the lowest computation time. Nevertheless, the *T-A* strategies inherit the limitations of the *T-A* formulation, i.e., they are intrinsically limited to cases where the 1D approximation of the superconducting layer is meaningful. On the contrary, the *H* strategies, except for the *H* homogenous strategy, are also suitable for the analysis of systems made of wires with different geometries, like MgB$_2$ wires.

It may be the case of a sufficiently large system where the memory recourses of the available computer are surpassed, even for the homogenous models. The *H* iterative multi-scale models have the additional advantage that they can be used to analyze large-scale systems almost without size limit. The reason is that the number of DOF of the single-tape model remains constant independently of the size of the system. The coil submodel, being a magnetostatic model, requires a lesser amount of computational resources to be simulated, therefore it is possible to analyze considerably larger systems.

# 8. References

[1] SuperPower Inc., "SuperPower," 2020. [Online]. Available: http://www.superpower-inc.com/. [Accessed: 10-Jan-2020].

[2] SuNam Co., Ltd, "SuNam," 2020. [Online]. Available: http://i-sunam.com/. [Accessed: 10-Jan-2020].

[3] Fujikura, "Fujikura," 2020. [Online]. Available: https://www.fujikura.com/solutions/superconductingwire/. [Accessed: 10-Jan-2020].

[4] B. G. Marchionini, Y. Yamada, L. Martini, and H. Ohsaki, "High-Temperature Superconductivity: A Roadmap for Electric Power Sector Applications, 2015-2030," *IEEE Trans. Appl. Supercond.*, vol. 27, no. 4, pp. 1–7, 2017.

[5] Z. Melhem, *High temperature superconductors (HTS) for energy applications*, 1st ed. Woodhead Publishing, 2011.

[6] S. S. Kalsi, *Applications of High Temperature Superconductors to Electric Power Equipment*, 1st ed. John Wiley and Sons Inc, 2011.

[7] H. Maeda and Y. Yanagisawa, "Recent Developments in High-Temperature Superconducting Magnet Technology (Review)," *IEEE Trans. Appl. Supercond.*, vol. 24, no. 3, pp. 1–12, 2014.

[8] Y. Lvovsky, E. W. Stautner, and T. Zhang, "Novel technologies and configurations of superconducting magnets for MRI," *Supercond. Sci. Technol.*, vol. 26, no. 9, p. 093001, 2013.

[9] R. Battesti *et al.*, "High magnetic fields for fundamental physics," *Phys. Rep.*, vol. 765–766, pp. 1–39, 2018.

[10] P. Komarek, "Advances in large scale applications of superconductors," *Supercond. Sci. Technol.*, vol. 13, no. 5, pp. 456–459, 2000.

[11] M. Zhang, W. Yuan, J. Kvitkovic, and S. Pamidi, "Total AC loss study of 2G HTS coils for fully HTS machine applications," *Supercond. Sci. Technol.*, vol. 28, no. 11, p. 115011, 2015.

[12] L. Quéval, V. M. R. Zermeño, and F. Grilli, "Numerical models for ac loss calculation in large-scale applications of HTS coated conductors," *Supercond. Sci. Technol.*, vol. 29, no. 2, p. 024007, 2016.

[13] F. Grilli, E. Pardo, A. Stenvall, D. N. Nguyen, W. Yuan, and F. Gomory, "Computation of losses in HTS under the action of varying magnetic fields and currents," *IEEE Trans. Appl. Supercond.*, vol. 24, no. 1, pp. 78–110, 2014.

[14] V. M. R. Zermeno, A. B. Abrahamsen, N. Mijatovic, B. B. Jensen, and M. P. Sørensen, "Calculation of alternating current losses in stacks and coils made of second generation high temperature superconducting tapes for large scale applications," *J. Appl. Phys.*, vol. 114, no. 17, p. 173901, 2013.

[15] A. Morandi, "2D electromagnetic modelling of superconductors," *Supercond. Sci. Technol.*, vol. 25, no. 10, p. 104003, 2012.

[16] Y. Yanagisawa, Y. Xu, X. Jin, H. Nakagome, and H. Maeda, "Reduction of screening current-induced magnetic field of REBCO coils by the use of multi-filamentary tapes," *IEEE Trans. Appl. Supercond.*, vol. 25, no. 3, pp. 1–5, 2015.

[17] Y. Li *et al.*, "Magnetization and screening current in an 800-MHz (18.8-T) REBCO NMR insert magnet: experimental results and numerical analysis," *Supercond. Sci. Technol.*, vol. 32, no. 10, p. 105007, 2019.

[18] W. T. Norris, "Calculation of hysteresis losses in hard superconductors carrying ac: Isolated conductors and edges of thin sheets," *J. Phys. D. Appl. Phys.*, vol. 3, no. 4, pp. 489–507, 1970.

[19] M. R. Halse, "AC face field losses in a type II superconductor," *J. Phys. D. Appl. Phys.*, vol. 3, no. 5, pp. 717–720, 1970.

[20] E. H. Brandt, "Thin superconductors in a perpendicular magnetic ac field: General formulation and strip geometry," *Phys. Rev. B*, vol. 49, no. 13, pp. 9024–9040, 1994.

[21] Y. Mawatari, "Critical state of periodically arranged superconducting-strip lines in perpendicular fields," *Phys. Rev. B - Condens. Matter Mater. Phys.*, vol. 54, no. 18, pp. 13215–13221, 1996.

[22] K. H. Müller, "Self-field hysteresis loss in periodically arranged superconducting strips," *Phys. C Supercond. its Appl.*, vol. 289, no. 1–2, pp. 123–130, 1997.

[23] J. R. Clem, "Field and current distributions and ac losses in a bifilar stack of superconducting strips," *Phys. Rev. B - Condens. Matter Mater. Phys.*, vol. 77, no. 13, p. 134506, 2008.

[24] G. P. Mikitik, Y. Mawatari, A. T. S. Wan, and F. Sirois, "Analytical methods and formulas for modeling high temperature superconductors," *IEEE Trans. Appl. Supercond.*, vol. 23, no. 2, pp. 8001920–8001920, 2013.

[25] F. Grilli, "Numerical modeling of HTS applications," *IEEE Trans. Appl. Supercond.*, vol. 26, no. 3, pp. 1–8, 2016.

[26] R. Brambilla, F. Grilli, L. Martini, and F. Sirois, "Integral equations for the current density in thin conductors and their solution by the finite-element method," *Supercond. Sci. Technol.*, vol. 21, no. 10, p. 105008, 2008.

[27] L. Prigozhin, "Analysis of critical-state problems in type-II superconductivity," *IEEE Trans. Appl. Supercond.*, vol. 7, no. 4, pp. 3866–3873, 1997.

[28] L. Prigozhin and V. Sokolovsky, "Computing AC losses in stacks of high-temperature superconducting tapes," *Supercond. Sci. Technol.*, vol. 24, no. 7, p. 075012, 2011.

[29] E. Pardo, F. Gömöry, J. Šouc, and J. M. Ceballos, "Current distribution and ac loss for a superconducting rectangular strip with in-phase alternating current and applied field,"

*Supercond. Sci. Technol.*, vol. 20, no. 4, pp. 351–364, 2007.

[30] E. Pardo, J. Šouc, and L. Frolek, "Electromagnetic modelling of superconductors with a smooth current-voltage relation: Variational principle and coils from a few turns to large magnets," *Supercond. Sci. Technol.*, vol. 24, no. 8, p. 044003, 2015.

[31] E. Pardo, "Modeling of screening currents in coated conductor magnets containing up to 40000 turns," *Supercond. Sci. Technol.*, vol. 29, no. 8, p. 085004, 2016.

[32] J. P. Bastos and N. Sadowski, *Electromagnetic Modeling by Finite Element Methods*, 1st ed. CRC Press, 2003.

[33] J.-M. Jin, *The Finite Element Method in Electromagnetics*, 2nd ed. New York: Wiley, 2002.

[34] J.-M. Jin, *Theory and Computation of Electromagnetic Field*. John Wiley & Sons, Inc., 2010.

[35] F. Grilli, R. Brambilla, and L. Martini, "Modeling high-temperature superconducting tapes by means of edge finite elements," *IEEE Trans. Appl. Supercond.*, vol. 17, no. 2, pp. 3155–3158, 2007.

[36] G. Barnes, M. McCulloch, and D. Dew-Hughes, "Computer modelling of type II superconductors in applications," *Supercond. Sci. Technol.*, vol. 12, no. 8, pp. 518–522, 1999.

[37] T. A. Coombs, A. M. Campbell, A. Murphy, and M. Emmens, "A fast algorithm for calculating the critical state in superconductors," *COMPEL - Int. J. Comput. Math. Electr. Electron. Eng.*, vol. 20, no. 1, pp. 240–252, 2001.

[38] A. M. Campbell, "A direct method for obtaining the critical state in two and three dimensions," *Supercond. Sci. Technol.*, vol. 22, no. 3, p. 034005, 2009.

[39] N. Amemiya, S. I. Murasawa, N. Banno, and K. Miyamoto, "Numerical modelings of superconducting wires for AC loss calculations," *Phys. C Supercond. its Appl.*, vol. 310, no. 1–4, pp. 16–29, 1998.

[40] A. Stenvall and T. Tarhasaari, "Programming finite element method based hysteresis loss computation software using non-linear superconductor resistivity and $T - \phi$ formulation," *Supercond. Sci. Technol.*, vol. 23, no. 7, p. 075010, 2010.

[41] R. Brambilla, F. Grilli, and L. Martini, "Development of an edge-element model for AC loss computation of high-temperature superconductors," *Supercond. Sci. Technol.*, vol. 20, no. 1, pp. 16–24, 2007.

[42] Z. Hong, A. M. Campbell, and T. A. Coombs, "Numerical solution of critical state in superconductivity by finite element software," *Supercond. Sci. Technol.*, vol. 19, no. 12, pp. 1246–1252, 2006.

[43] B. Shen, F. Grilli, and T. A. Coombs, "Review of the AC loss computation for HTS using H formulation," *Supercond. Sci. Technol.*, vol. 33, no. 3, p. 033002, 2020.

[44] H. Zhang, M. Zhang, and W. Yuan, "An efficient 3D finite element method model based on the T-A formulation for superconducting coated conductors," *Supercond. Sci. Technol.*, vol. 30, no. 2, p. 024005, 2016.

[45] F. Liang *et al.*, "A finite element model for simulating second generation high temperature

superconducting coils/stacks with large number of turns," *J. Appl. Phys.*, vol. 122, no. 4, p. 043903, 2017.

[46] E. Berrospe-Juarez, V. M. R. Zermeño, F. Trillaud, and F. Grilli, "Iterative multi-scale method for estimation of hysteresis losses and current density in large-scale HTS systems," *Supercond. Sci. Technol.*, vol. 31, no. 9, p. 095002, 2018.

[47] E. Berrospe-Juarez, V. M. R. Zermeño, F. Trillaud, and F. Grilli, "Real-time simulation of large-scale HTS systems: multi-scale and homogeneous models using the T–A formulation," *Supercond. Sci. Technol.*, vol. 32, no. 6, p. 065003, 2019.

[48] "COMSOL Multiphysics version 5.3." [Online]. Available: www.comsol.com. [Accessed: 10-Jan-2020].

[49] M. D. Ainslie, T. J. Flack, Z. Hong, and T. A. Coombs, "Comparison of first- and second-order 2D finite element models for calculating AC loss in high temperature superconductor coated conductors," *COMPEL - Int. J. Comput. Math. Electr. Electron. Eng.*, vol. 30, no. 2, pp. 762–774, 2011.

[50] J. Rhyner, "Magnetic properties and AC-losses of superconductors with power law current-voltage characteristics," *Phys. C Supercond. its Appl.*, vol. 212, no. 3–4, pp. 292–300, 1993.

[51] F. Sirois and F. Grilli, "Potential and limits of numerical modelling for supporting the development of HTS devices," *Supercond. Sci. Technol.*, vol. 28, no. 4, p. 043002, 2015.

[52] S. C. Chapra, *Métodos numéricos para ingenieros*, 5th ed. McGraw Hill, 2007.

[53] A. . Read, "Linear interpolation of histograms," *Nucl. Instruments Methods Phys. Res. Sect. A Accel. Spectrometers, Detect. Assoc. Equip.*, vol. 425, no. 1–2, pp. 357–360, Apr. 1999.

[54] W. D. Markiewicz *et al.*, "Design of a superconducting 32 T magnet with REBCO high field coils," *IEEE Trans. Appl. Supercond.*, vol. 22, no. 3, pp. 4300704–4300704, 2012.

[55] H. W. Weijers *et al.*, "Progress in the Development and Construction of a 32-T Superconducting Magnet," *IEEE Trans. Appl. Supercond.*, vol. 24, no. 3, pp. 1–5, 2016.

[56] L. Cavallucci, M. Breschi, P. L. Ribani, A. V. Gavrilin, H. W. Weijers, and P. D. Noyes, "A Numerical Study of Quench in the NHMFL 32 T Magnet," *IEEE Trans. Appl. Supercond.*, vol. 29, no. 5, pp. 1–5, 2019.

[57] E. Berrospe-Juarez, "Electromagnetic modeling of large-scale high-temperature superconductor systems," National Autonomous University of Mexico, 2020.

[58] E. Berrospe-Juarez, F. Trillaud, V. M. R. Zermeno, F. Grilli, H. W. Weijers, and M. D. Bird, "Screening Currents and Hysteresis Losses in the REBCO Insert of the 32 T All-Superconducting Magnet Using T-A Homogenous Model," *IEEE Trans. Appl. Supercond.*, vol. 30, no. 4, pp. 1–5, 2020.

[59] E. Berrospe-Juarez *et al.*, "Estimation of Losses in the (RE)BCO Two-Coil Insert of the NHMFL 32 T All-Superconducting Magnet," *IEEE Trans. Appl. Supercond.*, vol. 28, no. 3, pp. 1–5, 2018.